\documentclass{jpp-JAK}

\usepackage{bm,amsmath,amssymb,graphicx,array,url,natbib,slashed}
\usepackage[normalem]{ulem} % Wavy underlining, etc.
\usepackage[dvipsnames]{xcolor}
\usepackage{multicol}
\usepackage{upmath}

\usepackage[%dvips,% Doesn't work with \CJhapter, etc., as they stand
urlcolor={red!50!black}%
,citecolor={blue!80!black}%
,allbordercolors={0 0 0}%
,pdfborder={0 0 0}% Without braces, spaces seem to be stripped off.
,colorlinks=true%
,urlbordercolor={0 0 0}%
,hyperfootnotes=false%
,pdfborderstyle={/S/U/W 0}%
,plainpages=false%
,breaklinks=true%
]{hyperref}

\usepackage[all]{hypcap}

\usepackage{breakurl}

\usepackage{mathrsfs,pstricks}

\makeatletter

\gdef\@underjournal{%
  \vbox to 5.5\p@{\noindent
    \parbox[t]{4.5in}{\normalfont\indexsize{\itshape \
}\\[2.5\p@]
      {\ \ }}%
  \vss}%
}

\long\def\comment#1\endcomment{}

\let\marginparo\marginpar
\def\marginpar#1{\marginparo{\raggedright #1}}

\def\@sect
#1#2#3#4#5#6[#7]#8{%
\ifnum #2>\c@secnumdepth 
 \let \@svsec \@empty 
\else 
 \refstepcounter {#1}%
 \ifnum #2>\@ne 
  \protected@edef \@svsec {\@seccntformat {#1}\relax }%
 \else 
  \def \apphe@d {#8}%
  \protected@edef \@svsec {\@secappcntformat {#1}\relax }%
 \fi
\fi 
\@tempskipa #5\relax 
\ifdim \@tempskipa >\z@ 
 \begingroup 
 #6{\@hangfrom {\hskip #3\relax \@svsec }%
    \interlinepenalty \@M 
     #8\@@par 
   }
 \endgroup 
\csname #1mark\endcsname {#7}%
\ifnum #2=\@ne 
 \addcontentsline {toc}{#1}%
 {\ifnum #2>\c@secnumdepth \else 
   \ifappendix 
    \appendixname ~\csname the#1\endcsname:  #7% 
   \else 
    \protect \numberline {\csname the#1\endcsname .}#7%
   \fi 
  \fi 
  }%
\else 
 \addcontentsline {toc}{#1}%
 {\ifnum #2>\c@secnumdepth \else 
  \protect \numberline {\csname the#1\endcsname .}#7%
  \fi 
  }%
\fi 
\else 
 \def \@svsechd {#6{\hskip #3\relax \@svsec #8}%
 \csname #1mark\endcsname {#7}%
 \addcontentsline {toc}{#1}%
 {\ifnum #2>\c@secnumdepth \else 
  \protect \numberline {\csname the#1\endcsname .}%
  \fi #7%
  }}%
\fi 
\@xsect {#5}%
}

\let\cite\citep
\let\citeo\cite
\def\cite{\ \citeo}
\newcommand\Ref{\citet} % JPP

\catcode`\@=11 % Same as ``\makeatletter''.

\newif\ifjournal

\renewcommand*\boldmath{\protect\mathversion{bold}} % Add extra \protect
\newcommand*\Appabbrev{appendix} % The name to be used in text:  \App(A) ->
\newcommand*\Appsabbrev{appendixes}
\newcommand*\eq[1]{\label{#1}}	% Use inside an equation
\newcommand*\Eq[1]{Eq.~(\ref{#1})}
\newcommand*\Eqs[1]{Eqs.~(\ref{#1})}
\newcommand*\EQ[1]{\Unskip\EQo{#1}}
\newcommand*\Footnote[1]{Footnote~\seco{#1}}	% Start of sentence.

\comment
\newcommand*\Ref[1]{Ref.~\Onlinecite{#1}}

\newcommand*\REFo[1]{{\let\ \relax\Onlinecite{#1}}} % Can be used after a dash, like
\endcomment

\newcommand*\Onlinecite{\onlinecite} % For PRE, redefine to be surrounded
\newcommand*\seco[2][]{\ref{#2}#1}	
\newcommand*\Ifarrayflag{\ifarrayflag}
\newcommand*\Ifletterflag{\ifletterflag}
\newcommand*\Ifbeginningeq{\ifbeginningeq}

\newcommand*\BE{\arrayflagfalse\beginningeqfalse\begin{equation}} % *

\newcommand*\BEA{\arrayflagtrue\beginningeqfalse\begin{eqnarray}} % *
\newcommand*\BA{\BEA} %  \let; *
\def\BAams#1\EAams{\begin{align}#1\end{align}} % *

\newcommand*\EEA{\end{eqnarray}}
\newcommand*\BAL[1][]{\letterflagtrue\st@rtarray[#1]}
\newcommand*\EAL{\end{eqnarray}\EM}

\newcommand*\EE{%
\Ifbeginningeq
	\beginningeqfalse
	\BE
\else
	\Endanequation
	\beginningeqtrue
\fi
\Ifletterflag
	\EM
\fi
}

\newcommand*\Endanequation{%
\Ifarrayflag
	\end{eqnarray}%
\else
	\end{equation}%
\fi
}

\newcommand*\BM[1][]{\begin{subequations}%
	\gdef\theletters{a}%
	\def\NP##1{##1%
		\ifxf'@label\@empty\else\@xp\ltx@label\@xp{f'@label}\fi
		\letf'@label\@empty
		\def\theequation{\theparentequation\theletters}%
		\stepcounter{equation}%
		\protected@edef\@currentlabel{\theparentequation\alph{equation}}%
		\xdef\theletters{\theletters,\alph{equation}}%
		}%
	\def\NQ##1{\xdef\theletters{\alph{equation}}%
		\NP{##1}}%
	\l@belletters{#1}%
	}

\newcommand*\EM{\end{subequations}\COMMENT
	\letterflagfalse
	}

\newcommand*\NN{\nonumber}

\newcommand*\BI{\begin{itemize}}
\newcommand*\EI{\end{itemize}}

\newcommand*\verticalbar{|}
\newcommand*\eps{\varepsilon}

\newcommand*\ph{\varphi}

\newcommand*\s{\sigma}
\renewcommand*\th{\theta}

\newcommand*\w{\omega}

\newcommand*\shortGreek{
	\renewcommand*\a{\alpha}
	\renewcommand*\b{\beta}
	\renewcommand*\c{\chi}
	\renewcommand*\d{\delta}
	\newcommand*\D{\Delta}
	\renewcommand*\l{\lambda}
	\renewcommand*\L{\Lambda}
	\renewcommand*\t{\tau}
	\renewcommand*\u{\upsilon}
}

\newcommand*\rcvr@tmp[1]{\edef\next{\global\let\noexpand#1\csname#1@tmp}\next}

\newcommand*\longGreek{
	\rcvr@tmp{a}
	\rcvr@tmp{b}
	\rcvr@tmp{c}
	\rcvr@tmp{d}
	\rcvr@tmp{D}
	\rcvr@tmp{l}
	\rcvr@tmp{L}
	\rcvr@tmp{t}
	\rcvr@tmp{u}
}
	
\shortGreek

\newcommand*\BIGavg[1]{\left\langle#1\right\rangle}

\renewcommand*\({\left(}
\renewcommand*\){\right)}

\renewcommand*\[{\left[}

\newcommand*\?{}

\newcommand*\abso[1]{\verticalbar#1\verticalbar} % Absolute value, guaranteed to be small.

\newcommand*\At[1]{{}_{\big\vert#1}} % Sometimes a bigger bar looks better.
\newcommand*\at[1]{_{\vert#1}}

\renewcommand*\Bar[1]{{\overline{#1}}}	% Long over-bar.
\newcommand*\BL{Balescu--Lenard}

\newcommand*\bfit{\rmfamily\bfseries\itshape} % Boldfaced italic (really important stuff).
\newcommand*\bigast{{\mathchoice{\asterisk}% Displaystyle
	{\displaystyle\asterisk}% Textstyle
	{\textstyle\asterisk}% Scriptstyle
	{\scriptstyle\asterisk}% Scriptscriptstyle
	}}

\newcommand*\bo{b_0} 		% Minimum impact parameter
\newcommand*\conj{^\bigast}	% Complex conjugate.

 \newcommand*\Ekern{\kern-0.12em}
\newcommand*\degreeso{^\circ}		% The basic symbol for degrees.
\renewcommand*\degrees{\ifmmode\degreeso\else$\degreeso$\fi} % ATTN:  CONDITIONAL.

 \newcommand*\fkern{\kern-0.125em}
\newcommand*\diel{{\mathcal{D}}} 	% Full
\newcommand*\DiraC[1]{\delta\vlp#1\vrp}
\newcommand*\ed{\epsilon_\delta}
	
\newcommand*\ek{\v{\epsilon}_\vk}
\newcommand*\ETAL{\Latin{et~al.}}	% Use for journal (where \kill@period
\newcommand*\ETC{\LatinAIP{etc.}}		% Use for journal (where \kill@period
\newcommand*\half{\case12}		% Small 1/2.
\newcommand*\Int{\int\!\?}		% No limits
\newcommand*\I[2]{\int_{#1}^{#2}\!\?}	% General limits
\newcommand*\kD{k_{\Rm D}}		% Debye wave number
\newcommand*\khat{\unit{k}}		% Unit vector

\newcommand*\kill@period{\futurelet\nextchar\no@period}
\newcommand*\no@period{\ifx\nextchar.\skip@period\fi}	% ATTN:  CONDITIONAL.
\newcommand*\kmin{k\min}		% Minimum wave number.
\newcommand*\kmax{k\max}		% Maximum wave number.
\newcommand*\kpq[4]{#1_{\v{#2}\v{#3}\v{#4}}}
\newcommand*\Latin{\textit} % How to write Latin abbreviations such as ``ad~hoc''.
\newcommand*\LatinAIP{\textrm}	% AIP isn't consistent for more common ones.
\newcommand*\lD{\lambda_{\Rm D}}	% Debye length
\newcommand*\Mathop[1]{\mathop{\hbox{\rm #1}}\nolimits}
 \def\M(#1,#2,#3){\kpq{M}{#1}{#2}{#3}} % Mode-coupling coefficient (3 args)
\newcommand*\Max{\mathop{\operator@font max}}	% The maximum, as in \Max(1,2).
\renewcommand*\max{_{\textrm{max}}}
\newcommand*\Min{\mathop{\operator@font min}}	% The minimum.
\renewcommand*\min{_{\textrm{min}}}

\newcommand*\m[1]{^{-#1}}		% Raise to negative power.

\newcommand*\nbar{{\Bar{n}}}		% Mean
\newcommand*\on[1]{[#1]}			% Depends on (functionally).
\newcommand*\Rm[1]{{\rm #1}}			% The default for most journals.
\newcommand*\Schr{Schr\"odinger}
\newcommand*\sbar{{\Bar{s}}}

\newcommand*\set[1]{{\let|\mid \{#1\}}}	% Set (collection). (Redefine for
 \def\triad(#1,#2,#3){\kpq{\theta}{#1}{#2}{#3}} % Wave--wave interaction time.
\newcommand*\Unsym{^{\Rm U}}	% Unsymmetrized (mode-coupling coef.)
\newcommand*\Unsymconj{^{\Rm U\bigast}}	% Unsymmetrized (mode-coupling coef.) conj.
\newcommand*\unit[1]{{\widehat{\v{#1}}}}	% Unit vector.

\newcommand*\Vkern{\kern-0.1em} % Backspacing for subscripts of bold V's.

\renewcommand*\v[1]{{\protect\bm{#1}}} 	% Make a bold-faced vector.
\newcommand*\vk{\v{k}}
\newcommand*\vlp{\mathopen{\boldsymbol(}} % Boldfaced left paren.

\newcommand*\vrp{\mathclose{\boldsymbol)}} % Boldfaced right paren.

\newcommand*\vv{\v{v}}

\newcommand*\vvbar{{\Bar{\vv}}}

\newcommand*\Partial[1]{\Deriv\partial\fr{#1}}

\newcommand*\mathselect[4]{\mathchoice{#1{#3}{#4}}
	{#2{#3}{#4}}{#2{#3}{#4}}{#2{#3}{#4}}}

\newcommand*\frD{\frac} % \let
\newcommand*\frT[2]{#1/#2} % Set fraction in \textstyle

\newcommand*\fr[1]{\fro#1\fro} % Pull apart comma-delimited arguments.
 
\newcommand*\fR[1]{\left(\fr{#1}\right)}	
\newcommand*\Casefr[2]{\frac{#1}{#2}}	% A big (\displaystyle) fraction.

\newcommand*\Half{\Casefr12}	

\newcommand*\Fourth{\Casefr14}

	\renewcommand*\BE{\begin{equation}}
	\renewcommand*\EE{\end{equation}}
	\renewcommand*\BEA{\begin{eqnarray}}
	\renewcommand*\BAL[1][]{\BM[#1]\BA}
	\newcommand*\BALams[1][]{\BM[#1]\BAams}
\def\BALams{\@ifnextchar[\BALams@{\BALams@[]}}
\def\BALams@[#1]#2\EALams{\BM[#1]\BAams#2\EAams\EM}
	\renewcommand*\EM{\end{subequations}}
	\newcommand*\WT{\begin{widetext}}
	\newcommand*\NT{\end{widetext}}
	\renewcommand*\etal{\ETAL}	% \let; No \kill@period.
	\renewcommand*\etc{\ETC}	% \let; No \kill@period.
	\renewcommand*\fkern{\!}	% A bit too much backspace, but OK.
	\renewcommand*\Vkern{}		% No tiny backspace.
\newcommand*\l@belletters[1]{\def\Jtemp{#1}\ifx\Jtemp\empty\else\eq{#1}\fi}

 \newcommand*\activebar{\catcode`\|=\active}

 {\activebar
 \gdef\normalbar{\activebar
 	\let|\verticalbar}%
 }

 \normalbar

\def\Derivo#1#2#3,#4\Derivo{#2{#1{#3},#1#4}}

\def\Partial#1{\Deriv\partial\fr{#1}}

\def\mathselect#1#2#3#4{\mathchoice{#1{#3}{#4}}
	{#2{#3}{#4}}{#2{#3}{#4}}{#2{#3}{#4}}}

\def\fro#1,#2\fro{{\mathselect\frac\frT{#1}{#2}}}
\let\frD\frac
\def\frT#1#2{#1/#2} % Set fraction in \textstyle

\def\fr#1{\fro#1\fro} % Pull apart comma-delimited arguments.
\def\fR#1{\left(\fr{#1}\right)}	
\def\choiceo#1,#2\choiceo{{#1\atop#2}}

\def\Casefr#1#2{\frac{#1}{#2}}	% A big (\displaystyle) fraction.

\def\casefr#1#2{\mathchoice{{\textstyle\frD{#1}{#2}}}%
	{{\textstyle\frD{#1}{#2}}}% (\displaystyle; this is the odd one)
	{{\scriptstyle\frD{#1}{#2}}}%
	{{\scriptscriptstyle\frD{#1}{#2}}}}

\def\half{\casefr12}	\def\Half{\Casefr12}	
		
	\def\Fourth{\Casefr14}

\def\PARENS#1{\noexpand\let
\csname#1x\endcsname
\csname#1\endcsname
\noexpand\def
\csname#1(##1){\csname#1x\endcsname{##1}}}

\renewcommand*\l@belletters[1]{\def\Jtemp{#1}\ifx\Jtemp\empty\else\eq(#1)\fi}

 \let\COMMENT\relax % Used in the \EM macro above; overcomes a REVTeX
 \newif\ifarrayflag
 \newif\ifletterflag
 \newif\ifbeginningeq
 \beginningeqtrue

 \def\M(#1,#2,#3){\kpq{M}{#1}{#2}{#3}} % Mode-coupling coefficient (3 args)
 \def\Mpas(#1,#2,#3){\kpq{M\Unsym}{#1}{#2}{#3}} % As above, but passive.
 \def\Mpasconj(#1,#2,#3){\kpq{M\Unsymconj}{#1}{#2}{#3}} % As above, but passive.

\def\3{ ---\nobreak\ }

\renewcommand*\set[1]{{\let|\mid \{#1\}}}	% Set (collection).
\renewcommand*\etc{\ETC\kill@period}
\def\skip@period\fi.{\fi}

\newif\ifBIG

\def\SMALL{\BIGfalse}

\SMALL		% Keep things small by default.

\def\@BIGinsert#1{\ifBIG#1\fi}

\let\sc@le\empty

\def\SIZE#1{\edef\sc@le{\expandafter\noexpand\csname #1\endcsname}}

\def\MID{\mathbin{\hbox{\vrule height\ht0 depth\dp0 width0.2pt}}}

{\activebar
\gdef\<#1>{{\def\arg{\ifBIG\displaystyle\fi
\ifx\sc@le\empty
	\def\sc@lel{\@BIGinsert{\left}}
	\def\sc@ler{\@BIGinsert{\right}}
\else
	\let\sc@lel\sc@le
	\let\sc@ler\sc@le
\fi
\sc@lel
\langle
#1
\sc@ler
\rangle
}% End \def\arg
\let|\mid
\setbox0\hbox{$\arg$}
\ifBIG
	\let|\MID
\fi
\arg
}% End inner environment
\global\BIGfalse 	% Keep things small by default.
\global\let\sc@le\empty
}% End \gdef\<
}

\def\BIGavg#1>{\left\langle#1\right\rangle} % Grown brackets, middle bar
\def\bra#1|{\mathopen{}\ifBIG\left\fi\langle
	\,#1\,
	\ifBIG\BIGfalse\right\fi\verticalbar
	\mathclose{}}		% ATTN:  CONDITIONAL.

\def\ket#1>{\mathopen{}
	\ifBIG\left\fi\verticalbar
	\,#1\,
	\ifBIG\BIGfalse\right\fi\rangle
	\mathclose{}}		% ATTN:  CONDITIONAL.

\def\DOTS.{\unskip\ \dots\,.} % Used in quoted material when the last part
\renewcommand*\l@belletters[1]{\def\Jtemp{#1}\ifx\Jtemp\empty\else\eq{#1}\fi}

\renewcommand*\Eq[1]{\hyperref[#1]{Eq.}~(\ref{#1})}
\renewcommand*\Eqs[1]{\hyperref[#1]{Eqs.}~(\ref{#1})}
\def\FIGURE{\@ifnextchar[\@FIGURE{\@FIGURE[1]}}

\def\@FIGURE[#1]#2#3{\begin{figure}%[ht]
\vbox{%
\centerline{\includegraphics[width=#1\columnwidth]{#2.eps}}%
\nobreak
	    \caption{#3}%
	    \edef\@currentlabel{\thefigure}%
	    \label{Fg.#2}
}
\end{figure}}

\newcommand\footnoteref[1]{\protected@xdef\@thefnmark{\ref{#1}}\@footnotemark}

\renewcommand*\Partial[2]{\frac{\partial#1}{\partial#2}}

\renewcommand*\fr{\frac}
\renewcommand*\fR[2]{\left(\frac{#1}{#2}\right)}

\renewcommand*\[{\left[}

\let\asterisk*
\renewcommand\.{\boldsymbol{\cdot}}

\renewcommand*\At[2]{\mathord{\left.#1\right|\setbox0=\hbox{$\displaystyle#1$}%
\dimen0=\ht0
\advance\dimen0 by\dp0
\dimen0 = 0.3\dimen0
\!\lower\dimen0\hbox{$\scriptstyle#2$}}}

\renewcommand*\set[1]{\{#1\}}

\newcommand*\st{\Tilde s}

\let\chio\chi
\renewcommand*\chi{\Delta\chio}

\let\Delta\rmDelta

\newcommand*\Dskip[1]{\mskip#1.2mu}

\renewcommand*\M{{\rm M}}

\renewcommand*\P{{\rm P}}

\def\bra#1|{\mathopen{}\ifBIG\left\fi\langle
\Dskip{1}#1\Dskip{1}
	\ifBIG\BIGfalse\right\fi\verticalbar
	\mathclose{}}		% ATTN:  CONDITIONAL.

\def\ket#1>{\mathopen{}
	\ifBIG\left\fi\verticalbar
\Dskip{1}#1\Dskip{1}
	\ifBIG\BIGfalse\right\fi\rangle
	\mathclose{}}		% ATTN:  CONDITIONAL.

{\activebar
\gdef\<#1>{{\def\arg{\ifBIG\displaystyle\fi
\ifx\sc@le\empty
	\def\sc@lel{\@BIGinsert{\left}}
	\def\sc@ler{\@BIGinsert{\right}}
\else
	\let\sc@lel\sc@le
	\let\sc@ler\sc@le
\fi
\sc@lel
\langle
\Dskip{1}#1\Dskip{1}
\sc@ler
\rangle
}% End \def\arg
\def|{\Dskip{1}{\mid}\Dskip{1}}%
\setbox0\hbox{$\arg$}
\ifBIG
	\def|{\Dskip{5}{\MID}\Dskip{5}}%
\fi
\arg
}% End inner environment
\global\BIGfalse 	% Keep things small by default.
\global\let\sc@le\empty
}% End \gdef\<
}

\renewcommand\subsubsection{%
  \@startsection{subsubsection}{3}{\z@}
    {9\p@ \@plus 3\p@ \@minus 3\p@}
    {3\p@}
    {\raggedright\normalfont\normalsize}%
}

\def\HBOX#1{\hbox to 0pt{\hss #1\hss}}%

\def\subcenter#1{_{%
\dimen0 = 0pt % Holds maximum width of the \hbox's.
\toks0={}% Accumulates the \hbox's.
\def\Hbox##1{\hbox to \dimen0{\hss ##1\hss}}%
\vtop{\small\subparse#1{}}}}

\def\subparse#1{%
\def\next{#1}%
\ifx\next\empty
	\the\toks0 % Found end of list; spit out the boxes.
\else
	\setbox0=\hbox{\small#1}%
	\ifdim\wd0>\dimen0
		\dimen0=\wd0 % Calculate maximum width.
	\fi
	\toks0=\expandafter{\the\toks0\Hbox{#1}}
	\let\next\subparse
\fi
\next
}

\def\subnoparse#1{%
\def\next{#1}%
\ifx\next\empty
	\the\toks0 % Found end of list; spit out the boxes.
\else
	\setbox0=\hbox{\small#1}%
	\toks0=\expandafter{\the\toks0\Hbox{#1}}
	\let\next\subparse
\fi
\next
}

\def\Multiple#1#2#3{{\count0=#2
\toks0={#3}%
\loop\advance\count0 by -1
\ifnum\count0<0
\else
	\edef\temp{\toks0={#1{\the\toks0}}}%
	\temp
\repeat
\the\toks0}}

\newenvironment{Quote}
{\smallskip
\list{}{\leftmargin1.5\parindent\rightmargin\leftmargin\def\parsep{3pt}}\item\relax
\qsmall\baselineskip 10.75\p@
\noindent\ignorespaces
}
{\endlist\smallskip}

\def\precite#1#2{{\def\NAT@open{(#1; }\cite{#2}}}

\def\@hangfrom@appendix#1#2#3{#1\@if@empty{#2}{\MakeTextUppercase{#3}}{#2\@if@empty{#3}{}{:\ \MakeTextUppercase{#3}}}}
\def\@appendixcntformat#1{\MakeTextUppercase{\appendixname\ \csname
    the#1\endcsname}} 

\comment
\usepackage{xr}
\endcomment

\comment % PoP style

\renewcommand*\Eq[2][]{\setxdelim{#1}\hyperref[#1\xdelim#2]{Eq.}~(#1\xdelim\ref{#1\xdelim#2})} 
\renewcommand*\EQ[2][]{\setxdelim{#1}(#1\xdelim\ref{#1\xdelim#2})} 
\renewcommand*\Eqs[2][]{\setxdelim{#1}\hyperref[#1\xdelim#2]{Eqs.}~(#1\xdelim\ref{#1\xdelim#2})} 
\endcomment

\renewcommand*\Eq[2][]{\setxdelim{#1}(#1\xdelim\ref{#1\xdelim#2})} 
\renewcommand*\EQ[2][]{\setxdelim{#1}(#1\xdelim\ref{#1\xdelim#2})} 
\renewcommand*\Eqs[2][]{\setxdelim{#1}(#1\xdelim\ref{#1\xdelim#2})}

\newcommand*\nopdflink{\def\pdfmark[##1]##2{##1}}% Abort hyperlinking and
\newcommand\setxdelim[1]{\def\temp{#1}%
\ifx\temp\empty
 \def\xdelim{}%
\else
 \def\xdelim{:}%
 \nopdflink
\fi
}

\let\pio\pi
\renewcommand*\pi{\mathup\pio}
\let\Pi\pio

\newcommand\dd{\mbox{d}}

\newcommand\ii{{\rm i}}

\def\condition(#1){\hbox{if\ }#1}

\def\NAT@exlab{} % JPP

\def\Footnote#1{footnote~\ref{#1}%
\edef\curpage{\the\c@page}%
\ on page \pageref{#1}%
}
\def\pages#1#2{#1 (#2 pages)}
\def\range#1{\rangeo#1.}
\def\rangeo#1-#2.{#1-#2}

\def\killcomma#1{<KC>}

\long\def\pg#1#2#3{#1%
\ifx#2,%
 \ifx#3[%
  \ #3% No comma before [
 \else
  \ifx#3;%
   ; %
  \else
   #2 #3%
  \fi
 \fi
\else
 #2 #3%
\fi
}

\renewenvironment{Quote}
{
\smallskip
\list{}{\leftmargin1.25\parindent\rightmargin\leftmargin\def\parsep{0pt}\listparindent\parindent}\item\relax 
\small\noindent\ignorespaces
}
{\endlist\smallskip
}

\newcommand*\bmin{b_{\rm min}}
\newcommand*\bmax{b_{\rm max}}
\newcommand*\thmin{\theta_{\rm min}}
\newcommand*\tho{\theta_0}
\newcommand*\thB{\theta_B}
\renewcommand*\set[1]{\{#1\}}
\newcommand*\lBbar{\slashed{\lambda}_B}
\newcommand*\classical{_{\rm cl}}
\newcommand*\quantum{_{\rm qu}}

\newcommand\lB{\lambda_B}
\let\rperp\bo

\catcode`\@=12

\begin{document}

\title{An introduction to the physics of the Coulomb
  logarithm with emphasis on quantum-mechanical effects} 

\author{J. A. Krommes\corresp{\email{\protect\url{krommes@princeton.edu}}}}

\affiliation{Princeton Plasma Physics Laboratory, Princeton University, MS
  28, P.O. Box 451, Princeton, NJ  08543--0451 USA}

\maketitle

\begin{abstract}
An introduction to the physical interpretation of the Coulomb logarithm is
given with particular emphasis on the quantum-mechanical corrections that
are required at high temperatures.  Excerpts from the literature are used
to emphasize the historical understanding of the topic, which emerged more
than a half-century ago.  Several misinterpretations are noted.
Quantum-mechanical effects are related to diffraction by scales of the
order of the Debye screening length; they are not due to quantum
uncertainty related to the much smaller distance of closest approach.
\end{abstract}

% The following redefine the footnoting scheme to be numerical (approved by
% Alex Schekochihin.
{\makeatletter

\global\let\@makempfntext\@makempfntexto
%\global\let\@makefntext\@makefntexto
\global\let\@mpfootnotetext\@mpfootnotetexto
\global\let\footnoterule\footnoteruleo
\global\let\@makefnmark\@makefnmarko
\global\let\thempfootnote\thempfootnoteo
\global\let\thefootnote\thefootnoteo

\global\let\@fnsymbol\@fnsymbolo

\global\def\@makefntext#1{\strut $^{\the\c@footnote}$#1}

%\global\renewcommand\@makempfntext[1]{\strut $^{\the\c@footnote}$#1}
\global\setlength\footnotesep{10\p@}
}

% Table of contents
\setcounter{tocdepth}{3} % Default is evidently 2
{
\def\boldmath{}
\def\bfit{}
\tableofcontents
}

\section{Introduction}

The Coulomb logarithm $\ln\Lambda$, often called the `Spitzer logarithm'
in honor of 
its discussion in the pioneering monograph of \Ref{Spitzer} (and earlier
by \Ref*{Cohen} in the section prepared by
Spitzer) is one of 
the most fundamental quantities in basic plasma physics, as it quantifies the
dominance of small-angle scattering in a weakly coupled plasma.
\comment
Recently, Mulser \etal\cite{Mulser} strongly criticized Spitzer's physical
interpretation of the quantum-mechanical correction to~$\Lambda$.  However,
as I shall describe, a review of Spitzer's discussion and related literature
reveals that Mulser \etal\ 
misunderstood Spitzer's argument, and that the interpretation by
Spitzer,\cite{Spitzer,Cohen} Sivukhin,\cite{Sivukhin}
Kulsrud,\cite{Kulsrud} Wesson,\cite{Wesson} 
and others is correct.
\endcomment
In introductory discussions of collisions in plasmas, usually only
classical physics is considered.  However, for sufficiently high
temperatures quantum-mechanical effects become important.  Although the
proper way of including those effects has been understood at least since
the work of Cohen \etal, there appears to be some ignorance of that early
literature and, thus, some residual uncertainty about the
interpretation of the quantum-mechanical corrections.  The purpose 
of this brief tutorial is to provide a new student of the subject with some
entry points to the literature relating to the basic
physics of the Coulomb logarithm.
Sometimes-lengthy excerpts from original papers and reviews are used to
emphasize the 
historical development.  Several interesting points of confusion are identified.

\comment
In fact, the ultimate conclusion
obtained by Mulser \etal\ actually agrees with that of the authors cited in the
previous paragraph.  However, the description of Spitzer's argument given
by Mulser \etal\ adds unnecessary confusion to a most 
important topic.  The purpose of my comments below is to clarify Spitzer's
arguments and to supplement the
final points made by Mulser \etal\ by displaying some excerpts from the
original literature.  The
discussion is intended to provide 
\endcomment

I shall address only the restricted problem of the calculation and
interpretation of~$\ln\L$ in weakly coupled, many-body, charge-neutral
plasmas with the 
usual statistical symmetries of homogeneity and isotropy; other issues,
such as the effects of anisotropy considered in the interesting work by
\Ref*{Mulser}, are not discussed. 

\section{The Coulomb logarithm in classical kinetic theory}

One definition of~$\L$ is $\L \doteq \lD/\bmin$, where $\lD$~is the Debye
screening length (the effective maximum impact parameter for two-body
scattering) and $\bmin$~is a 
characteristic length to be determined.  In classical scattering theory,
 $\bmin = \bo$, where 
\BE
\bo \doteq \fr{q_1 q_2}{\mu u^2}
\EE
is the impact parameter for $90\degrees$ scattering between particles~1
and~2.  (Here $q$~is the charge, $\mu$~is the reduced mass, and $u$~is the
relative velocity; $\doteq$ denotes a definition.  The distance of closest
approach is~$2\bo$.) This result arises in
classical, weakly coupled plasma kinetic theory\cite{Montgomery1} as
follows.  One can calculate the net scattering cross section~$\s$, or in
more detail the velocity-dependent collision operator, by an appropriate
integration over all possible impact parameters~$b$ or, equivalently, by an
integration over wave-number magnitude $k \doteq b\m1$.
In particular, with $\ek \doteq
-4\pi\ii \vk/k^2$ being the Fourier representation of the bare Coulomb
field of a unit point charge and $\diel(\vk,\w)$ being the Vlasov
dielectric function, 
the \BL\ collision operator (which captures the effect of dielectric
shielding but not large-angle scattering) is\footnote{The derivation of the
  \BL\ operator  
  is reviewed in section~G.1 of \Ref{JAK_projection_II}.}
\BALams
C_{s\sbar}\on{f} &\doteq -\pi(\nbar m)_s\m1(\nbar q^2)_s (\nbar q^2)_{\sbar}
\Partial{}{\vv}\.\Int \dd\vvbar\Int \fr{\dd\vk}{(2\pi)^3}
\fr{\ek\,\ek\conj}{\abso{\diel(\vk,\vk\.\vv)}^2}\DiraC{\vk\.(\vv - \vvbar)}
\NN\\
&\qquad\qquad\.\(\fr{1}{m_s}\Partial{}{\vv} -
\fr{1}{m_{\sbar}}\Partial{}{\vvbar}\)
f_s(\vv)f_{\sbar}(\vvbar)
\eq{C_BL}
\\
&\sim 
\I0{\kmax}\fr{dk}{k}\fR{1}{[1 + A(k\lD)\m2]^2 + [B(k\lD)\m2]^2},
\eq{I_BL}
\EALams
where $A(\khat,\vv)$ and~$B(\khat,\vv)$ are functions of order unity.  The
denominator inside the large parentheses represents the mean-square effect
of dielectric 
shielding and makes the integral strongly convergent at $k \to 0$,
effectively limiting the wave-number integration to $k \gtrsim \kmin \doteq
\kD$,  where $\kD \doteq \lD\m1$; this corresponds to a maximum impact
parameter $\bmax \approx \lD$.  (The 
integral \EQ{I_BL} can be done exactly, but the details are not important
for qualitative discussion.)
Because Debye
shielding is ineffective at small scales,
\Eq{I_BL} is logarithmically divergent at large~$k$ and must be
cut off at some~$\kmax$ (or~$\bmin$). However, this divergence is not
physical.  It 
arises because the \BL\ derivation is
perturbative, based on zeroth-order straight-line trajectories; thus, large-angle
scattering (the effect of impact parameters $b\lesssim\bo$) is
misrepresented.  There is no solution for this difficulty within the \BL\
framework.  However, one can asymptotically
match between an `inner' solution that treats large-angle scattering and
an `outer' solution appropriate for small-angle scattering, as was done, for
example,  
by \Ref{Frieman-Book} for the classical regime.  Such
matching removes any apparent divergence at the small scales, in agreement
with the result that the Rutherford cross section is integrable as the
scattering angle $\th \to 
\pi$ or $b \to 0$.  Of course, the scale~$\bo$ remains in the
asymptotically matched solution, as it determines the two-body relationship
between scattering angle and impact parameter, 
\BE
\tan(\th/2) = \bo/b,
\eq{tan}
\EE
and sets
the basic size of the classical cross section~$\s\classical$, which is finite.
If the convergent integral that defines~$\s\classical$ is then approximated by
ignoring dielectric shielding, one finds, as a consequence of the
long-ranged nature of the Coulomb force, that
the integral $\I0{\bmax} db\ldots \sim \I{}{\bmax}db/b$
exhibits a logarithmic divergence at
\emph{large}~$b$ that must be rectified by a cutoff at~$\bmax$.
That introduces the Coulomb logarithm;
the result is dominantly $\s\classical \sim \bo^2\ln\L\classical$,
where
\BE
\ln\L_{\rm cl} \equiv L_{\rm cl} \doteq \ln(\lD/\bo).
\eq{L_cl}
\EE
When this same approximation is applied to \Eq{C_BL}, one is led to
  the Landau collision operator\cite{Landau36}, which is generally used in
  practice.  The 
  Landau operator is discussed in appendix~B of \Ref{JAK_projection_I}.

An alternative way of arriving at $\ln\L\classical$ is to
begin with the Boltzmann collision operator (which does not
incorporate Debye shielding but does handle large-angle scattering
correctly), then to take
the small-angle limit.  That introduces 
the integral
$\I{\kmin}{\kmax}dk\,k\m1 = \I{\bmin}{\bmax}db\,b\m1$, to be considered in
the classical limit. In order to deal with the misrepresentation of
large-angle scattering in lieu of complicated asymptotic 
analysis,  students are generally taught to merely insert the 
short-scale cutoff $\bmin = \bo$ or $\kmax=\bo\m1$, which describes the
crossover between large- and small-angle scattering.  In any such
discussion, it should be stressed that that recipe results from
a systematic asymptotic matching between the classical regimes of large and
small impact parameters, that the underlying physics is convergent for small
impact parameters, and that no
discontinuity or exponentially rapid variation that would lead to
diffraction occurs in the vicinity of~$\bo$.  These remarks expand upon
some of those in the paper by \Ref{Mulser}.

When the impact-parameter integration is
transformed to an integration over scattering angle~$\th$, one sees that
the effect of dielectric shielding is to introduce a cutoff at
\emph{small}~$\th$, so $\s \sim 
\I{\thmin}{}d\th/\th \sim \ln\thmin\m1$; see, for example,
\Ref[Eq.~(41.9)]{Landau_PK}. Classically, one has $\thmin \approx \tho \doteq
2\bo/\lD$ (which follows from the small-angle limit of \Eq{tan}). In the
next section, we shall see that quantum-mechanical 
diffraction effects lead to a larger value of~$\thmin$ for sufficiently
high temperatures, and thus to a smaller value of~$\ln\L$.

\section{\boldmath Quantum-mechanical corrections to
  \texorpdfstring{$\ln\L$}{ln Lambda}}

As noted by Spitzer and more clearly  
explained in the 149-page review article by \Ref{Sivukhin}, at
sufficiently 
high temperature the classical approximation fails, quantum-mechanical
considerations apply, and one finds 
$\bmin =\lB \doteq  h/\mu u$,  the `de~Broglie wavelength of the test
particle in the coordinate system in which the scattering center \dots\ is
at rest'\cite{Sivukhin}.  Interpolation between the classical and
quantum-mechanical limits leads to the standard prescription 
\BE
\bmin = \Max\set{\lB,\bo}.
\eq{max}
\EE

\comment
This recipe was already known to Spitzer.  However, Mulser \etal\
questioned Spitzer's heuristic explanation of the appearance of~$\lB$ and
even questioned the validity of \Eq{max} itself.
Mulser \etal\ stated,
\begin{Quote}
``A special argument for [the prescription \EQ{max}] is by
Spitzer.\cite{Spitzer} 
He arrives 
at the limitation $b \ge \lB$ by observing that for impact parameters $b
\le \lB$ the Coulomb differential cross section leads to higher diffraction
values than an opaque disc of the same radius, which is `unphysical.'  It
seems that for numerous researchers this constitutes the basic argument.
\dots\
Although physically 
appealing at first glance, it is false and self contradictory.  In the
neighborhood of the Coulomb singularity, the author compares Rutherford
scattering with optical diffraction from a diaphragm of diameter $2\lB$.
\dots\
Spitzer's setting of $b_{\min} = \lB$ is a prominent
example of excellent physical intuition but mistaken proof.''
\end{Quote}
They also asserted, ``there is no basis for such a rule as $\bmin =
\Max\set{\lB,\bo}$. It is a mere guess, no proof has ever  been given.''

Examination of Spitzer's discussion 
and related literature cited below belies these assertions of Mulser
\etal\  In 
fact, 
\endcomment
\noindent
\Ref[p.~128]{Spitzer} said it this way:\footnote{To conform
  to my notation, I have changed Spitzer's~$p$ to~$b$, and $w$~to~$u$.} 
\begin{Quote}
When the electron temperature exceeds about $4\times 10^5$ degrees~K,
$\Lambda$~must be somewhat reduced below the values obtained from
[classical theory], because of quantum-mechanical effects.  An electron
wave passing 
through a circular aperture of radius~$b$ will be spread out by diffraction
through an angle $\l/2\pi b$, where $\l$~is the electron wavelength.  If
this deflection exceeds the classical deflection $2\bo/b$, then the
previous equations must be modified; in accordance with the results of
\Ref{Marshak} the only change needed is to reduce~$\L$ by the ratio $\a
c/u$, where $\a$~is the fine structure constant, equal to $1/137$.
\end{Quote}
Essentially the same discussion appears in \Ref[p.~233]{Cohen}.
\comment
Nowhere in this argument did Spitzer mention ``a diaphragm of diameter
$2\lB$'' or discuss impact parameters smaller than~$\lB$.
\endcomment
Spitzer 
referred to the choice of a deflection angle, called~$\thmin$ above.
His words correspond to the fact that when 
$\lB > \bo$ diffraction of the de~Broglie wave by an opaque disc of
radius~$\lD$
(not the very much smaller radius~$\lB$)
produces a
diffraction angle~$\thB$ that is larger than the 
classical deflection angle~$\tho$ for a particle incident
with maximum impact parameter $\bmax = \lD$; thus, $\s\quantum \sim \ln\thB\m1 <
\s\classical \sim \ln\tho\m1$. 

A possible source of confusion is that Spitzer did not completely spell out
the argument; he did not explicitly state that the diffraction angle
should be evaluated with $\bmax = \lD$ although this is clearly implied by
the fact that 
$\lD$~appears in the classical~$\tho$, and by Spitzer's discussion (on the
page preceding the above quotation)
of~$\lD$ as the maximum impact parameter).  Instead, he cited a paper by
\Ref{Marshak}.  Tracing back through the historical record is
interesting.  Marshak stated,\footnote{To conform to my notation, I
  have replaced Marshak's~$\tho$ by~$\thmin$.}
\begin{Quote}
Thus far we have not given any explicit form for~$I$ [$I$~being the
$\th$~integral of the Rutherford differential scattering cross
section~$\s(\th)$].  If we look at the 
integral expression for~$I$ [\Eq{5.10} below with $\eps = 0$] we see that it
diverges if we integrate between the limits~0 and~$\pi$.  However, there are
physical grounds for extending this integration only to some small
angle~$\thmin$, in which case:
$$
2I = \Mathop{log}_e\fr{2}{(1 - \cos\thmin)}.
$$
Now it can be shown that $\thmin = \fr{\slashed{\lambda}}{a}$\,* where
$\slashed{\lambda}$ is the de~Broglie wavelength of the electron
participating in the collision, and ``$a$'' is the screening radius
\dots\,.
\end{Quote}
Marshak referred to the screening radius `of the atom', but it is clear
that in the 
present plasma context one should replace~$a$ by~$\lD$.  Thus, we see more
clearly 
that Spitzer was recapitulating and providing a physical interpretation of
Marshak's argument, which was focused on the 
physics of~$\thmin$ or, equivalently, the physics associated with maximum
impact parameter~$\lD$.  In turn, Marshak's *~footnote, which provides the
background to his assertion `it can be shown that \dots', refers to p.~497
of the famous paper of 
\Ref{Bethe33} on the quantum mechanics of one- and two-electron
atoms.  Bethe worked in the first Born approximation (for which he cited
earlier literature) and discussed a form factor~$F$ that determines~$\s(\th)$.
In modern notation, the formula for~$\s(\th)$ is given by \Eq{5.8}.
(This reduces to the Rutherford cross section when $\thB\to0$.)
Bethe emphasized,\footnote{In English:  
`The atomic form factor~$F$ itself
  depends on 
  the scattering angle~$\th$ --- more precisely on $\sin(\th/2)/\l$.'}
\begin{Quote}
Der Atomformfaktor~$F$ selbst h\"angt vom
Ablenkungswinkel~$\th$ --- genauer von $\fr{\sin\th/2}{\lB}$ ---~ab.
\end{Quote}
(The last ratio is not dimensionally correct; obviously, the intended
comparison is between~$\sin\half\th$ and~$\half\thB$ or, for small angle,
between~$\th$ and~$\thB$.)  It is the quantum correction to the Rutherford
cross section, which is important at small scattering angle and was well
known to the pioneers of quantum mechanics, that 
underlies Marshak's conclusion and Spitzer's interpretation.  

The first Born approximation does not, in and of itself, predict the
interpolation recipe \EQ{max}. 
Indeed, in that approximation the integral that defines the total momentum
transfer 
can be done exactly, a result known to Sivukhin [see \Eq{5.8} in the 
excerpt below] and surely earlier workers as well.  
The inner length~$\bo$ enters the resulting formula only
multiplicatively (the total scattering cross section is proportional
to~$\bo^2$). However, the first 
Born approximation is valid (at best; see later discussion) only at high
energies. Sivukhin explains the 
issue clearly:\footnote{To conform to my notation, I have changed
  Sivukhin's~$D$ to~$\lD$.}
%(apparently Mulser \etal\ were not aware of Sivukhin's article):
\begin{Quote}
3.  The classical-mechanics analysis applies so long as $(2\pi/\l)\rperp \gg
1$, where $\l = h/\mu u$ is the de~Broglie wavelength of the test particle
in the coordinate system in which the scattering center (field particle)
is at rest.  \dots\ We can write this condition in the
form
\BE
u \ll \a c,
\eq{5.4}
\EE
where
\BE
\a = \left|\fr{e\,e\conj}{\hbar c}\right|
\eq{5.5}
\EE
($\hbar = h/2\pi = 1.05\cdot 10^{-27} \hbox{erg}\cdot\hbox{sec}$).  If~$e$
and~$e\conj$ are equal to the elementary charge the constant $\a =
e^2/\hbar c = 1/137$ is the fine-structure constant.

The classical analysis cannot be used if \Eq{5.4} is not satisfied.  This
result might appear strange at first glance since the exact
quantum-mechanical solution for scattering of a charged particle in a
Coulomb field yields an expression for $\s(\th,u)$ which is exactly the
same as the classical expression \dots\ (cf., for example,
\Ref{Davydov} or
any text on quantum mechanics).  The essential point here, however, is that the
results coincide only when the scattering field is a Coulomb field over all
space.  In the case of a cutoff Coulomb field the wave properties of the
particle are appreciably different from those given by the classical
analysis.

When
\BE
u \gg \a c,
\eq{5.6}
\EE
the quantum-mechanical scattering problem can be solved relatively easily
by means of the Born approximation.  The solution of the problem is
simplified if the cutoff Coulomb field is replaced by a Debye field with
potential
\BE
\ph = \fr{e\conj}{r}e^{-r/\lD}.
\eq{5.7}
\EE 
If \Eq{5.6} is satisfied, it is well known that the quantum-mechanical
analysis leads to the result
\BE
\s(\th,u) = \fR{e\,e\conj}{2\mu u^2}^2\fr{1}{\(\sin^2\fr{\th}{2} +
  \eps^2\)^2},
\eq{5.8}
\EE
where
\BE
\eps = \fr{\l}{4\pi \lD} = \fr{\hbar}{2\mu u \lD}
\EE
(cf., for example, [any textbook discussion of quantum-mechanical
scattering theory]). 

By substituting \EQ{5.8} [into the formula for mean momentum transfer] we
recover [the classical result \EQ{L_cl}] with the sole difference that the
classical value of the Coulomb 
logarithm is replaced by the quantum-mechanical value
\BAams
L_{\rm qu} &= \Fourth\I0\pi \fr{\sin^2\fr{\th}{2}
    \sin\th}{\(\sin^2\fr{\th}{2} + \eps^2\)^2}\,d\th
\NN\\
&= \Half\ln\fr{1 +
    \eps^2}{\eps^2} - \fr{1}{2(1 + \eps^2)}.
\eq{5.10}
\EAams

In all cases of physical interest it is found that
\BE
\eps = \fr{\l}{2\pi \lD} \ll 1,
\EE
so that the square of~$\eps$ can be neglected compared with unity.  In this
approximation
\BE
L_{\rm qu} = \ln\fr{1}{\eps} - \Half = \ln\fr{4\pi \lD}{\l} - \Half.
\eq{5.12}
\EE
If the term $-1/2$ is neglected this expression differs from the classical
value [$\ln(\lD/\bo)$] only in that the lower limit~$\rperp$ is replaced
by~$\l/4\pi$.  This result is easily understood:  The De Broglie wave
associated with the incident particle is diffracted on the Debye sphere
surrounding the scattering center.  Diffraction theory or elementary
interference considerations shows that to within a factor of order unity
the mean value of the diffraction angle is $\th = \l/2\lD$.  If this value
exceeds the classical limit $\th_{\rm min} = 2\rperp/\lD$, the classical
formula \dots\ no longer applies and $\th = \l/2\lD$ is to be taken as the
lower limit in the integral in Eq.~(3.8).  This procedure leads to $L =
\ln(4\lD/\l)$, which differs from \Eq{5.12} by the unimportant factor
of~$\pi$ under the logarithm.  It also follows from this qualitative
description that scattering on a cutoff Coulomb field is essentially the
same as scattering on a Debye field \EQ{5.7}.

The quantum-mechanical relation \EQ{5.12} can be written in the form
\BE
L_{\rm qu} = L_{\rm cl} + \ln\fr{2\a c}{u} - \Half
\eq{5.13}
\EE
[which shows that when the diffraction correction is valid the size of the
Coulomb logarithm is reduced].

We recall that this formula is derived under the assumption that $u \gg \a
c$, whereas the classical expression \EQ{L_cl} applies when $u \ll \a c$.

\smallskip

\noindent
4.  The quantum-mechanical calculation becomes extremely complicated in the
intermediate region.  It would not be very meaningful to be concerned with
this region because we are already using the binary-collision approximation
with its artificial and, indeed, somewhat arbitrary truncation of the
Coulomb forces so that any improvement in the values of the Coulomb
logarithm obtained as a result of more complex calculations would be quite
illusory.  Instead, it is simpler and more in the spirit of the
approximation used here to proceed as follows.  The formulas for the
limiting cases \EQ{L_cl} and \EQ{5.12} show that the Coulomb logarithm
contains the velocity~$u$ under the logarithm so that the latter is a
slowly varying function of~$u$.  It is physically obvious that this slow
variation also obtains in the intermediate region.  Hence, without
incurring any serious error we can extrapolate \Eqs{L_cl} and \EQ{5.12} into
the intermediate region up to the value $u = u_{\lim}$, at which both
expressions coincide.  When $u < u_{\lim}$ the classical formula
\EQ{L_cl} is to be used; when $u \gg u_{\lim}$ the quantum-mechanical
formulas \EQ{5.12} or \EQ{5.13} are used.

\dots If the relative velocity~$u$ is replaced by the equivalent temperature
according to the relation $3T = \mu u^2$ \dots\ [and upon]
substituting the appropriate values of the reduced mass for a deuterium
plasma, we obtain the following limiting temperatures for
electron--electron, electron--ion, and ion--ion collisions, respectively:
\BE
\begin{cases}
T_{\lim}^{ee} = 6.65\,\hbox{eV},\\
T_{\lim}^{ei} = 13.3\,\hbox{eV},\\
T_{\lim}^{ii} = 2.45\cdot 10^4\,\hbox{eV} = 24.5\,\hbox{keV}.
\end{cases}
\eq{Tlims}
\EE
\end{Quote}
A similar discussion can be found on p.~239 of the book by
\Ref{Kulsrud}, who cites Sivukhin.  (Kulsrud arrives at
somewhat different but qualitatively similar transition temperatures by
using a different interpolation scheme.)

All of the above discussions are consistent in their physical
interpretations.  Interestingly, however, \Ref{Mulser} challenged Spitzer's
heuristic description.  They stated,
\begin{Quote}
A special argument for [the prescription \EQ{max}] is by
\Ref{Spitzer}.  He arrives 
at the limitation $b \ge \lB$ by observing that for impact parameters $b
\le \lB$ the Coulomb differential cross section leads to higher diffraction
values than an opaque disc of the same radius, which is `unphysical.'  It
seems that for numerous researchers this constitutes the basic argument.
\dots\
Although physically 
appealing at first glance, it is false and self contradictory.  In the
neighborhood of the Coulomb singularity, the author compares Rutherford
scattering with optical diffraction from a diaphragm of diameter $2\lB$.
\dots\
Spitzer's setting of $b_{\min} = \lB$ is a prominent
example of excellent physical intuition but mistaken proof.
\end{Quote}
This characterization of Spitzer's discussion is incorrect and
 appears to be a misunderstanding; nowhere in his argument did
Spitzer mention `a diaphragm of diameter 
$2\lB$' or discuss impact parameters smaller than~$\lB$.  In fact,
\comment
They also asserted, ``there is no basis for such a rule as $\bmin =
\Max\set{\lB,\bo}$. It is a mere guess, no proof has ever  been given.''
\endcomment
in agreement with the various authors cited above, Mulser \etal\ also
concluded that 
the de~Broglie correction was associated with large impact parameters.
But although they
asserted that this was a new and surprising result, we see that
it has been understood for more than a half-century.  

Although the basic ideas and results are clear, some discussions in the
literature are incomplete; for example, 
Mulser \etal\ cited a number of references in which apparently the classical
cutoff was used.  However, closer inspection shows that in several of the
papers cited by Mulser \etal\ 
the authors were, in 
fact, aware of the quantum correction.  For example, \Ref*{RMJ} refer in
their footnotes~6 and~3 to the discussion of 
cutoffs by \Ref{Cohen} which as we have seen contains the
original version of Spitzer's argument.  And although
\Ref{Balescu_transport_1} described the classical
situation, he noted (p.~130), 
\begin{Quote}
Some 
authors (e.g.\
Braginskii (1965) consider semi-empirical corrections to $\ln\Lambda$ under
various conditions of temperature and density; we do not discuss these
minor points here.
\end{Quote}
One could quibble with Balescu's
characterization of the issue as `minor', and the first
Born approximation does not deserve to be called `semi-empirical'.
In any event, it is instructive to consider the explanation of
\Ref[p.~238]{Braginskii}, who said, 
\begin{Quote}
At large
velocities, in which case $e^2/hv < 1$, where $h$~is Planck's constant
(i.e., $v/c < 1/137)$, it is necessary to use a smaller value for the
maximum impact parameter; specifically, we use the distance for which the
scattering angle is of the same order as the quantum uncertainty, in which
case $p_{\rm max} \approx \d_D e^2/hv$.
\end{Quote}
Here it is asserted that instead of using the classical formula $\thmin \sim
\bo/b_{\rm max}$ with $b_{\rm max} = \lD$, one should use $b_{\rm max} =
\lD(\bo/\lB)$ or  $\thmin \sim \lB/\lD$, the latter `quantum uncertainty'
being the 
diffraction angle of a 
wave with wavelength~$\lB$ encountering an object of radius~$\lD$.
But although Braginskii's argument leads to the correct~$\thmin$,
his heuristic introduction of a modified~$b_{\rm max}$ is incorrect;
apparently it 
was devised in order to obtain agreement with the proper~$\thmin$
 and the earlier discussion of Spitzer.  However, Braginskii did appreciate the
 role of 
 the quantum-mechanical uncertainty due to diffraction by Debye screening
 clouds of radius~$\lD$, and he understood
 that the
classical cutoff was not to be used for large velocities.

\comment
Mulser \etal\ suggested that `many'
researchers share the incorrect view that the appearance of~$\lB$
in~$\ln\L$ arises from quantum uncertainty at the small $\bo$~scales.  This
is questionable in view of the clear explanation in Sivukhin's important
and widely known review,
but nevertheless it is useful to dispel any residual confusion.  (The following
remarks expand on those by Mulser \etal\ in their second-to-last paragraph.)
Such confusion could arise when one considers
the scattering problem as formulated in terms of incoming impact
parameter~$b$ instead of outgoing
scattering angle~$\th$, as above.
The impact-parameter representation often arises in treatments
of weakly coupled plasma kinetic theory, where it is typical to represent

?????
\endcomment

Ultimately, the proper value of~$\thmin$ must follow from a systematic
kinetic theory. Clearly, $\lB$~can enter the problem
only when 
quantum-mechanical effects are included.  An asymptotic matching between
the Born approximation and the quasi-classical regime is described by
\Ref[\S46]{Landau_PK}, where original references are
cited.
%\footnote{Recent calculations by R.~Kulsrud (private communication,
%  2018) are in essential agreement with those of Landau and Lifshitz in
%  Ref.~\onlinecite{Landau_PK}.} 
The physics content of that calculation agrees with the heuristic arguments
given above. 

\comment
And of course one should
point out the heuristic de~Broglie
modification to the classical result, which
encapsulates a more rigorous asymptotic matching that includes
quantum-mechanical effects\cite{Landau_PK}.
\endcomment

A final topic concerns the validity of the first Born approximation.
\comment
Mulser \etal\ argued that the approximation holds provided that the ratio $r
\doteq a \bo/\lBbar$ is small, where $\lBbar \doteq \lB/2\pi$ and $a
\approx 0.2$.  That estimate is difficult, but it differs by merely a
numerical coefficient from the 
ratio used by Sivukhin (see the first sentence of the excerpt above).
Mulser \etal\ observe that for a tokamak $r \approx 10\m2$ (for electron--ion
scattering, this corresponds to an effective temperature of
$3.6\,\hbox{keV}$), thus conclude that the first Born approximation is
valid.  This agrees with Sivukhin's conclusions about transition
temperatures [\Eq{Tlims}] based on 
interpolating $\ln\L$ between the classical and quantum regimes. (Note
that for such temperatures ion--ion collisions should be represented
classically.)  
However, although the first Born approximation leads to sensible results,
it is not rigorously valid.  
\endcomment
Kulsrud\footnote{R. Kulsrud, private communication (2018).} has considered
the issue in detail and concluded that the first Born approximation is not
rigorously valid even when 
$r\doteq a \bo/\lBbar$ is small.
He points out that criteria for the validity of that
approximation, such as given by \Ref{Schiff}, are not satisfied for
the calculation of the Coulomb logarithm; `a full solution of the \Schr\
equation is necessary.'  However, his more exact analysis, similar in
spirit to that 
described by \Ref{Landau_PK}, corrects the simpler
estimates described 
above by a relatively small amount in the five per cent range.

\section{Summary and discussion}
In summary,
excerpts from the
literature provide historical
perspective and explain the correct interpretation of the
quantum-mechanical correction to $\ln\Lambda$.
The~$\lB$ in the ratio $\L\quantum = \lD/\lB$ that appears in the first Born
approximation (which is, roughly speaking, valid for sufficiently high
temperatures) 
`is the result of the charge distributions at large impact 
parameters'\cite{Mulser}, not quantum uncertainty related to the impact
parameter~$\bo$ for 90\degrees\ scattering or to the distance of closest
approach ($2\bo$).  This conclusion was obtained early on 
 by \Ref{Cohen}, \Ref{Spitzer},
and \Ref{Sivukhin}, and it has been usefully repeated in more modern texts such
as those of 
\Ref{Kulsrud} and \Ref{Wesson}.  Proper understanding of the physics of the
Coulomb logarithm is crucial, as that basic quantity figures in a multitude of
important plasma-physics applications.

\begin{acknowledgments}
I am grateful to Russell Kulsrud for informative discussions, and to Greg
Hammett and Ed Startsev for critical comments on early drafts of the manuscript.
This work was supported by the U. S. Department of Energy Contract
DE-AC02-09CH11466.
\end{acknowledgments}

\bibliographystyle{jpp-JAK}

%\bibliography{/u/krommes/Tex/jak}

\end{document}